\documentclass[12pt]{article}
\usepackage {psfrag}
\usepackage{epsfig}
\usepackage{graphics}
\newcommand {\beq}{\begin{equation}}
\newcommand {\eeq}{\end{equation}}
\newcommand {\beqy}{\begin{eqnarray}}
\newcommand {\eeqy}{\end{eqnarray}}
\begin{document}
\begin{center}
{\bf Spin dynamics of a one-dimensional spin-$\frac{1}{2}$
fully anisotropic Ising-like antiferromagnet in a transverse
magnetic field}
\vskip 1cm

Asimkumar Ghosh

Department of Physics, Scottish Church College,
1 \& 3 Urquhart Square,\\ Kolkata 700 006, India.
\end{center}
\begin{abstract}
We consider the one-dimensional Ising-like fully anisotropic
S=$\frac{1}{2}$ Heisenberg antiferromagnetic Hamiltonian and study
the dynamics of domain wall excitations in the presence of
transverse magnetic field $h_x$. We obtain dynamical spin
correlation functions along the magnetic field $S^{xx}(q,\omega)$
and perpendicular to it  $S^{yy}(q,\omega)$. It is shown that the
line shapes of $S^{xx}(q,\omega)$ and  $S^{yy}(q,\omega)$ are
purely symmetric at the zone-boundary. It 
is observed in $S^{yy}(q,\omega)$ for $\pi/2<q<\pi$ that 
the spectral weight moves toward low energy side with the
increase of $h_x$. This model is applicable to study the spin
dynamics of CsCoCl$_3$ in the presence of weak
interchain interactions.
\end{abstract}
\section{Introduction}
The spin-$\frac{1}{2}$  Ising-like antiferromagnetic (AFM)
 chain has been the subject of theoretical studies for
 quite some time. The spin dynamics of the system are
 characterized by a picture of propagating domain walls or
 solitons. The magnetic compounds CsCoCl$_3$ and CsCoBr$_3$
are good examples of $S\!=\!\frac{1}{2}$ Ising-like AFM chains.
The simplest exchange interaction Hamiltonian describing these
compounds is the $S\!=\!\frac{1}{2}$ XXZ Heisenberg model
\beq
H_{\rm XXZ}=2J\sum_i\left[ S_i^zS_{i+1}^z+
\epsilon \left(S_i^xS_{i+1}^x+ S_i^yS_{i+1}^y\right)\right],
\;\;\;\;\;\;0<\epsilon<1.
\label{eq1}
\eeq
For very small $\epsilon$, the lowest order ground state of
Eq.(\ref{eq1}) is the N\'eel states with a $z$-component of the
total spin given by $S^z_{\rm T}$=0. Villain \cite{Villain}
has calculated the longitudinal correlation function
$S^{zz}(q,\omega)$ based on the basis states consisting of
a single domain wall and predicted the appearance of a central
peak with sharp shoulders. On the other hand, Ishimura and Shiba
\cite{IS} proposed a picture of domain wall pair (DWP) states
and showed that the propagating DWPs give rise to an excitation
continuum around the Ising excitation energy 2$J$. The transverse
correlation function $S^{xx}(q,\omega)$ exhibits a broad
peak around 2$J$. The existence of these peaks of
$S^{zz}(q,\omega)$ and $S^{xx}(q,\omega)$ has been verified
by inelastic neutron scattering experiments on
CsCoCl$_3$ \cite{Cscocl1,Cscocl2,Cscocl3} and CsCoBr$_3$ \cite{Cscobr}. 
 A significant feature of the spin wave
response of $S^{xx}(q,\omega)$ near the zone centre ($q\!=\!\pi$)
is that the spectral weights are heavily concentrated
towards the lower energy region. Nagler $et.\, al.$ \cite{Cscobr}
added a staggered field term
\beq
H_{\rm S}=h\sum_i(-1)^iS_i^z
\label{eq2}
\eeq
to the Hamiltonian in Eq.(\ref{eq1}). The staggered field $h$ has
two contributions $h_o$ and $h_{\rm ic}$. The first contribution
originates from taking account of the exchange mixing of higher
levels with the ground doublet. The second contribution arises
from the interchain exchange interactions at low temperatures.
The interchain interactions treated in the mean-field
approximation, give rise to the staggered field term $h_{\rm ic}$.
The effective Hamiltonian contains both the terms
$H_{\rm XXZ}$ and $H_{\rm S}$. With this effective Hamiltonian,
the broad peak is found to spilt into discrete peaks which is
known as Zeeman ladder and observed in Raman scattering on
CsCoCl$_3$ and CsCoBr$_3$ \cite{Lehmann}. However, the obsereved line shapes
of $S^{xx}(q,\omega)$ are quite different from those
of the theoretical predictions. Matsubara $et.\,al.$
\cite{Matsubara} have included a weak next nearest neighbour
(NNN) ferromagnetic (FM) interaction $H_{\rm F}$ in the
Hamiltonian $H_{\rm XXZ}$ in Eq.(\ref{eq1}):
\beq
H_{\rm F}=-2J^\prime\sum_i\left[ S_i^zS_{i+2}^z+
\epsilon \left(S_i^xS_{i+2}^x+ S_i^yS_{i+2}^y\right)\right].
\eeq
They have shown the existence of bound states of DWPs as well
as the free DWP states and the transverse correlation function
 $S^{xx}(q,\omega)$ exhibits a sharp peak at lower energy region.
The effect of transverse magnetic field $h_x$ on the
spin dynamics of this model has been studied by Murao $et.\,al.$
\cite{Murao} and shown that the spectral weight  moves towards
the low energy side in $S^{yy}(q,\omega)$ for $\pi/2<q<\pi$
with the increase in $h_x$, while there is no appreciable change
in $S^{xx}(q,\omega)$ for all $q$. The distribution of
intensities of the sharp peaks in $S^{yy}(q,\omega)$
vary irregularly for $q\approx \pi$. Although the proposed form
of NNN FM coupling provides a good description of most of
the experimental results, the required magnitudes of the NNN
exchange $|J^\prime|\sim 0.1 |J|$ is unphysically large
\cite{Goff}. In 1996, Bose and Ghosh \cite{Indrani}
have proposed the Ising-like fully anisotropic Heisenberg AFM
Hamiltonian in 1D and shown that the asymmetric line shapes of
$S^{xx}(q,\omega)$ and the bound states of DWPs can be derived.

In the absence of the magnetic field, $S^z_{\rm T}$ is a good
quantum number and the eigenvalues of different $S^z_{\rm T}$
having unequal number of DWPs form different energy bands
separating by energy 2$J$. In the presence of longitudinal
magnetic field, $h_z$, $S^z_{\rm T}$ still is a good quantum
number and the eigenvalues within the same value of $S^z_{\rm T}$
as well as the position of the peak of $S^{xx}(q,\omega)$
shift parallel with the increase in $h_z$. However,
$S^z_{\rm T}$  is no longer a good quantum number in
presence of a transverse magnetic field $h_x$ and a mixing of
states with different $S^z_{\rm T}$ occurs. Thus,
eigenvalues as well as eigenstates will be modified by $h_x$
and the characteristics of the spin dynamics  
will be different.

In this paper, we study the effect of $h_x$ on the dynamical spin 
correlation functions 
in a fully anisotropic Ising-like $S\!=\!\frac{1}{2}$ Heisenberg
AFM chain at low temperatures. Dynamical correlation functions
 $S^{xx}(q,\omega)$  and $S^{yy}(q,\omega)$ have been derived
 using the picture of propagating DWPs. Finally we introduce
 this model to explain the spin dynamics of CsCoCl$_3$ taking into account the
 weak interchain interactions (Eq.(\ref{eq2})). In Section 2,
 the theory and the results for the eigenvalues of the DWP
 continuum and DWP bound states are derived. The dynamical
 spin correlation functions of CsCoCl$_3$ are presented in
 Section 3. Section 4 contains a discussion of the
 results obtained.
 \section{Model and domain wall pair states}
 The one dimensional fully anisotropic Ising-like
 Heisenberg Hamiltonian in the presence of transverse
 magnetic field is given by
 \beqy
H&=&2\sum_i^N\left[ J_x S_i^xS_{i+1}^x+
J_y S_i^yS_{i+1}^y+ J_zS_i^zS_{i+1}^z\right]
-g_\perp\mu_{\rm B}H_x\sum_i^N S_i^x \nonumber \\
&=&2J\sum_i^N \left[ S_i^zS_{i+1}^z+\frac{\epsilon_1}{2}
\left( S_i^+S_{i+1}^-+S_i^-S_{i+1}^+\right)+
\frac{\epsilon_2}{2}
\left( S_i^+S_{i+1}^++S_i^-S_{i+1}^-\right)
\right]-h_x\sum_i^NS_i^x \nonumber \\
&&J=J_z,\;\;\epsilon_1=\frac{J_x+J_y}{2J},\;\;
\epsilon_2=\frac{J_x-J_y}{2J},\;\;h_x=g_\perp\mu_{\rm B}H_x,\;\;
\epsilon_1,\;\epsilon_2\ll1.
\label{hxyz}
\eeqy
$H_x$ is the transverse magnetic field and assume $h_x\ll 2J$. $N$
is the total number of spins.
Since we are interested in excitations at low temperatures, 
 we consider low lying excited states. These states
can be obtained from the N\'eel state by flipping a block of
adjacent spins, giving rise to DWP states with
$S_{\rm T}^z$=0 and $\pm$1 (Fig. 1). 
These excitations occur around the
Ising energy $2J$ above the ground state. Following the
method introduced by Murao $et.\,al.$ \cite{Murao},
we classify these states into two series. Series `a' starts from
the state with  $S_{\rm T}^z$=1 where two domain walls
are adjacent. Let $m$ be the number (odd) of sites between two
domain walls and $\phi^{(a)}_1(m)$ be the corresponding Ising
state. The subsequent states $\phi^{(a)}_j(m)\,(j=2,\,3,\,4,\,
 \cdot \,\cdot\, \cdot)$ are generated from $\phi^{(a)}_1(m)$
 such that the separation between the domain walls is increased by
 unit lattice distance successively towards the
 right hand side of the chain. Hence,
\beqy
 \phi^{(a)}_1(m)&=&S_m^+\,|{\rm N{\acute e}el}\rangle\;\;\;\;\;\;\;\;
 \quad\quad\quad\quad\quad\quad \quad \quad
 \quad\quad\quad \quad S_{\rm T}^z=1\nonumber \\
 \phi^{(a)}_j(m)&=&S_{m+j-1}^- \phi^{(a)}_{j-1}(m),\;\;
 (j=2,\,4,\,6,\,\cdot\,\cdot\,\cdot)\;\;\; \quad \quad  S_{\rm T}^z=0 \\
\phi^{(a)}_j(m)&=&S_{m+j-1}^+ \phi^{(a)}_{j-1}(m),\;\;
 (j=3,\,5,\,7,\,\cdot\,\cdot\,\cdot)\;\;\; \quad \quad S_{\rm T}^z=1  \nonumber
\label{eqa}
\eeqy
\begin{figure}[t]
\begin{center}
\psfrag{n1}{$|{\rm N{\acute e}el1}\rangle$}
\psfrag{n2}{$|{\rm N{\acute e}el2}\rangle$}
\psfrag{pa1}{$\phi_1^{(a)}(1)$}
\psfrag{pa2}{$\phi_2^{(a)}(1)$}
\psfrag{pa3}{$\phi_3^{(a)}(1)$}
\psfrag{pb1}{$\phi_1^{(b)}(1)$}
\psfrag{pb2}{$\phi_2^{(b)}(1)$}
\psfrag{pb3}{$\phi_3^{(b)}(1)$}
\includegraphics{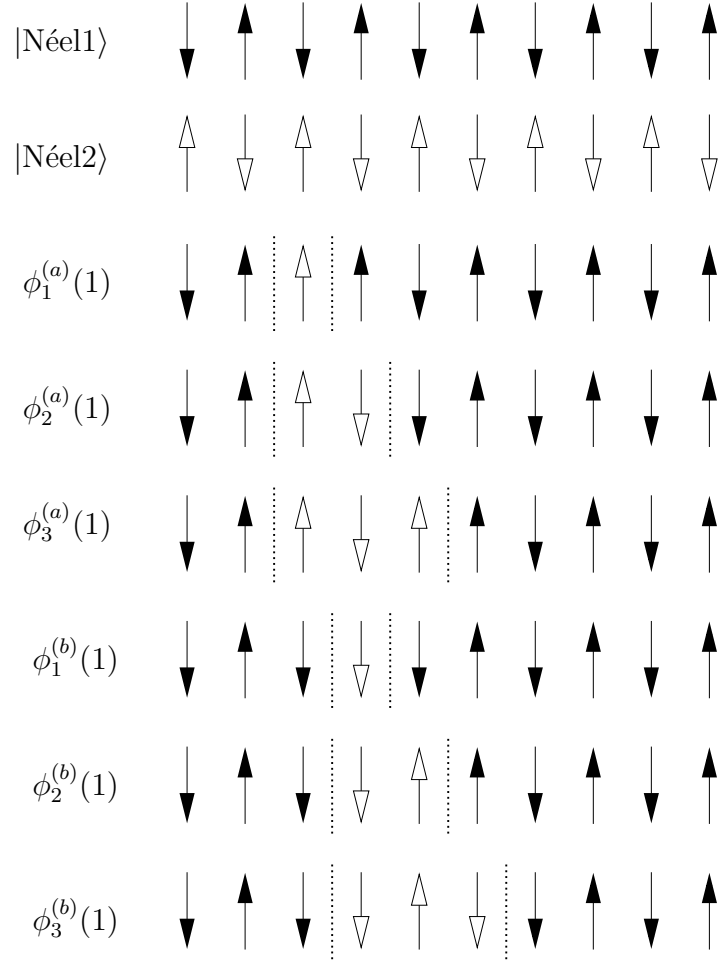}
\label{states}
\caption{N\'eel states and DWP states for $S_{\rm T}^z=\pm 1 \;{\rm and}\; 0$. 
The dotted vertical lines indicate the position of domain walls.}
\end{center}
\end{figure}
$ |{\rm N{\acute e}el}\rangle$ is one of the N\'eel states.
We choose a linear combination of these basis states  for describing 
propagating DWPs with wave vector $q$ as
\beq
|j,q\rangle_a=\sqrt{\frac{2}{N}}\sum_{m={\rm odd}}e^{-iqm} \phi^{(a)}_j(m)
\eeq
On the other hand, series `b' originates from the state with
$S^z_{\rm T}=-1$, and the subsequent states with
$S^z_{\rm T}$=0 and $-1$ appear alternately,
\beqy
 \phi^{(b)}_1(m)&=&S_m^-\,|{\rm N{\acute e}el}\rangle\;\;\;\;\;\;\;\;
 \quad\quad\quad\quad\quad\quad \quad \quad
 \quad\quad\quad \quad S_{\rm T}^z=-1,\nonumber \\
 \phi^{(b)}_j(m)&=&S_{m+j-1}^+ \phi^{(b)}_{j-1}(m),\;\;
 (j=2,\,4,\,6,\,\cdot\,\cdot\,\cdot)\;\;\; \quad \quad  S_{\rm T}^z=0, \\
\phi^{(b)}_j(m)&=&S_{m+j-1}^- \phi^{(b)}_{j-1}(m),\;\;
 (j=3,\,5,\,7,\,\cdot\,\cdot\,\cdot)\;\;\; \quad \quad S_{\rm T}^z=-1.  \nonumber
\eeqy
Taking linear combination of these states with wave vector $q$
\beq
|j,q\rangle_b=\sqrt{\frac{2}{N}}\sum_{m={\rm even}}e^{-iqm} \phi^{(b)}_j(m)
\label{eqb}
\eeq
With the help of the Eqs.(5)-(8), one can obtain
$H|j,q\rangle_a$ as:
\beqy
H|1,q\rangle_a &=&2J|1,q\rangle_a+V_{\epsilon_1}|3,q\rangle_a+
V_{\epsilon_2}|1,q\rangle_b   -\frac{h_x}{2}
\left(|2,q\rangle_a +e^{-iq}|2,q\rangle_b\right)\nonumber \\
H|2,q\rangle_a &=&2J|2,q\rangle_a+V_{\epsilon_1}|4,q\rangle_a
-\frac{h_x}{2}\left(|1,q\rangle_a + |3,q\rangle_a +e^{-iq}|3,q\rangle_b
+e^{iq}|1,q\rangle_b \right) \nonumber\\
\vdots && \vdots  \nonumber \\
H|j,q\rangle_a &=&2J|j,q\rangle_a+V_{\epsilon_1}|j\!+\!2,q\rangle_a+
V^*_{\epsilon_1}|j\!-\!2,q\rangle_a   -\frac{h_x}{2}
\left(|j\!-\!1,q\rangle_a + |j\!+\!1,q\rangle_a \right) \nonumber \\
&&-\frac{h_x}{2}\left(e^{iq}|j\!-\!1,q\rangle_b +
e^{-iq}|j\!+\!1,q\rangle_b\right),\;\;\;\;\;\;\;\;j\geq 3,\nonumber \\
\eeqy
where $V_{\epsilon_1}=\epsilon_1 J\left( 1+e^{-2iq}\right)$ 
 and  $V_{\epsilon_2}=2 \epsilon_2 J\cos q$.  In the same manner,
 one could derive similar set of equations for  $H|j,q\rangle_b$ in
 terms of $|n,q\rangle_a$ and $|n,q\rangle_b$. To avoid the mixing
  between the states of series `a' and `b', we further
 introduce symmetric ($\alpha$) and antisymmetric ($\beta$)
 functions \cite{Murao} defined as
\beqy
|j,q\rangle_\alpha &=&\frac{1}{\sqrt 2}\left( |j,q\rangle_a+|j,q\rangle_b\right),\nonumber \\
|j,q\rangle_\beta &=&\frac{1}{\sqrt 2}\left( |j,q\rangle_a-|j,q\rangle_b\right).
 \eeqy
 Hence, one can express $H|j,q\rangle_\alpha$ as
\beqy
H|1,q\rangle_\alpha &=&(2J+V_{\epsilon_2})|1,q\rangle_\alpha
+V_{\epsilon_1}|3,q\rangle_\alpha
  - V_\alpha |2,q\rangle_\alpha,  \nonumber \\
H|2,q\rangle_\alpha &=&2J|2,q\rangle_\alpha +V_{\epsilon_1}|4,q\rangle_\alpha
-\left(V^*_\alpha |1,q\rangle_\alpha +V_\alpha |3,q\rangle_\alpha  \right), \nonumber\\
\vdots && \vdots  \nonumber \\
H|j,q\rangle_\alpha &=&2J|j,q\rangle_\alpha+V_{\epsilon_1}|j\!+\!2,q\rangle_\alpha+
V^*_{\epsilon_1}|j\!-\!2,q\rangle_\alpha   -\left(V^*_\alpha |j\!-\!1,q\rangle_\alpha + V_\alpha |j\!+\!1,q\rangle_\alpha \right)  \nonumber
\\
&&\;\;\;\;\;\;\;\;\;\;\;\;\;\;\;\;\;\;\;\;\;\;\;\;\;\;\;\;\;
\;\;\;\;\;\;\;\;\;\;\;\;\;\;\;\;\;
\;\;\;\;\;\;\;\;\;\;\;\;\;\;\;\;
\;\;\;\;\;\;\;\;\;\;\;\;\;\;\;\;\;\;\;\;\;j\geq 3,
\label{eqal}
\eeqy
where $V_\alpha=\frac{h_x}{2}\left( 1+e^{-iq}\right)$. Similarly,
one can derive $H|j,q\rangle_\beta$ with $\alpha$ and $V_\alpha$ being
replaced by $\beta$ and
$V_\beta=\frac{h_x}{2}\left( 1-e^{-iq}\right)$, respectively. The
first excited states can be constructed as a linear combination
of symmetric and antisymmetric functions separately,
\beq
\Psi_\alpha(q)=\sum_j\alpha_j|j,q\rangle_\alpha \;\;\;{\rm and} \;\;\;
 \Psi_\beta(q)=\sum_j\beta_j|j,q\rangle_\beta  .
\label{eqeif}
\eeq
With the help of the Eq.(\ref{eqal}), the following equations
for the coefficients $\alpha_j$ and $\beta_j$ are obtained as
\beqy
\lambda_\alpha \bar \alpha_1&=&(2J+V_{\epsilon_2})\bar \alpha_1+
\bar V_\alpha \bar \alpha_2+\bar V_{\epsilon_1}\bar \alpha_3,
\nonumber \\
 \lambda_\alpha \bar \alpha_2&=&2J\bar \alpha_2+
\bar V_\alpha \left(\bar \alpha_1+\bar \alpha_3\right)
+\bar V_{\epsilon_1}\bar \alpha_4,
\nonumber \\
\vdots &&\vdots \nonumber \\
\lambda_\alpha \bar \alpha_j&=& 2J\bar \alpha_j+
\bar V_\alpha \left( \bar \alpha_{j\!-\!1}+\bar \alpha_{j\!+\!1}\right)
+\bar V_{\epsilon_1}\left(\bar \alpha_{j-2}+\bar \alpha_{j+2}\right),\;\;\;j\geq 3, 
\label{eqeig}
\eeqy
where $\lambda_\alpha$ is the eigenvalue,
$\bar V=-h_x\cos{\left(\frac{q}{2}\right)},\,\bar V_{\epsilon_1}=
2\epsilon_1J\cos q\;{\rm and} \;\bar \alpha_j=
\alpha_je^{\frac{iq}{2}j}$. In the same manner, one can derive
similar set of equations for $\beta$ with $\alpha$
being replaced by $\beta,\,\bar V_\alpha \,{\rm by} \,\bar V_\beta=h_x
\sin{\left(\frac{q}{2}\right)}\;{\rm and} \;\bar \alpha_j\;
{\rm by}\;\bar \beta_j=\beta_je^{\frac{i(k+\pi)}{2}j}$.

\begin{figure}[t]
\begin{center}
\psfrag{q}{$q$}
\psfrag{l1}{$\lambda_\alpha/J$}
\psfrag{l2}{$\lambda_\beta/J$}
\psfrag{h0}{$h_x\!=\!0$}
\psfrag{h1}{$h_x\!=\!0.1J$} 
\psfrag{p/2}{$\pi/2$}
\psfrag{p}{$\pi$}
\includegraphics{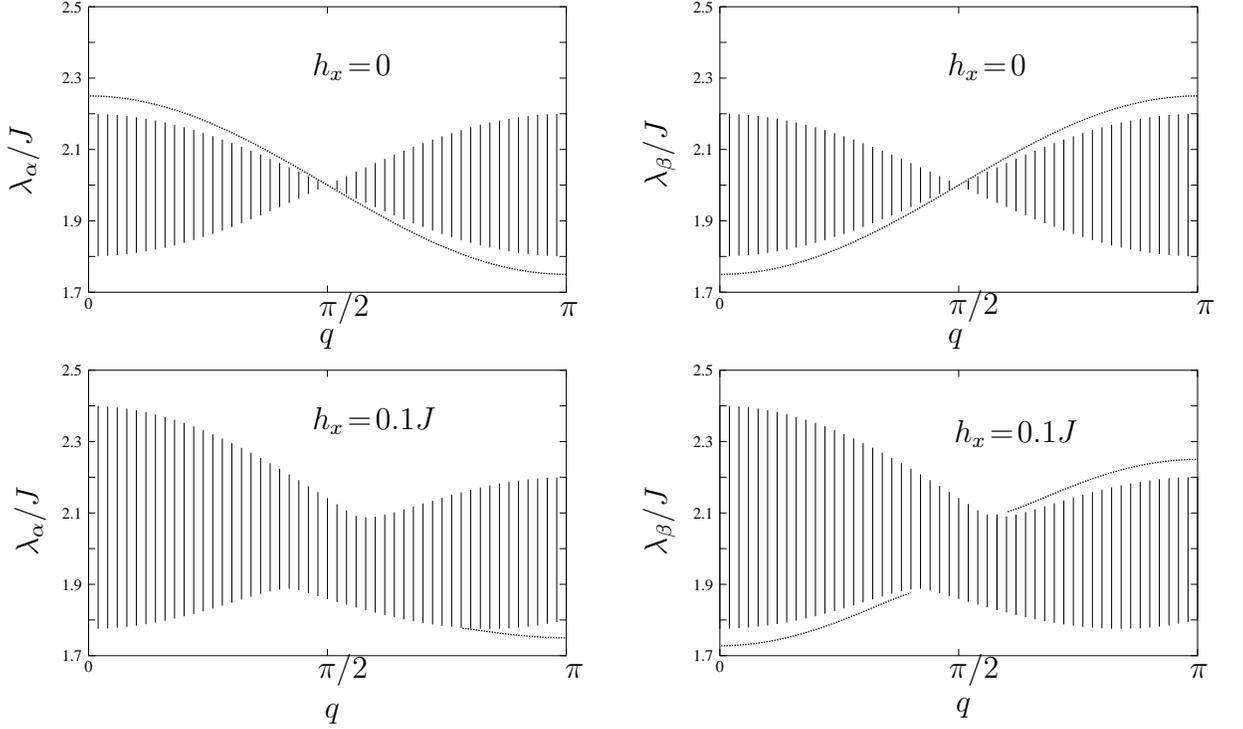}
\label{fig2}
\caption{Spin-wave excitation continuum (solid lines) and DWP 
bound-state energies (dotted lines) of 
the symmetric and antisymmetric modes for 
$\epsilon_1\!=\!0.05$ and $\epsilon_2\!=\!0.1$.}
\end{center}
\end{figure}
Dispersion relations are obtained numerically by solving
Eqs.(\ref{eqeig}), with N=1000. Here, we present the results
for $\epsilon_1=0.05$ and $\epsilon_2=0.10$, since these
values are estimated in the compound CsCoCl$_3$ \cite{Indrani}.
Figure 2 shows the disperson relations in the
symmetric and antisymmetric modes, $\lambda_\alpha$ and
$\lambda_\beta$, respectively,  for $h_x=0$. The spin wave continuum
and the bound state energy are plotted by solid and dotted
lines, respectively. In the symmetric mode,
the bound state energy lies above the continuum
for $q<\frac{\pi}{2}$ and below the continuum for
$q>\frac{\pi}{2}$, while the reverse is true for the
antisymmetric mode. The bound state does not exist
when $\epsilon_1>\epsilon_2$. When $h_x\neq 0$, the energy
band extends towards the high energy region for
$0\leq q\leq \frac{\pi}{2}$  in both the symmetric and the antisymmetric 
modes. The spin wave excitations have a width at $q=\frac{\pi}{2}$
in contrast with the case of $h_x=0$. The width also
broadens with the increase of $h_x$. Note that the
bound state energy is not affected by the presence of $h_x$.

\section{Dynamical spin correlation functions at T=0 K}
The dynamical spin correlation functions along the
direction of $h_x$ at T=0 is defined as
\beq
S^{xx}(q,\omega)=\sum_{\rm e}\left|\left\langle \Psi_{\rm e}\left|
S^x(q)\right|\Psi_{\rm g}\right\rangle\right|^2
\delta\left( \omega-\lambda_{\rm e}+\lambda_{\rm g}\right)
\label{eqsxx}
\eeq
 where $|\Psi_{\rm g}\rangle$, $|\Psi_{\rm e}\rangle$
 denote the ground and excited states, respectively, and
 $\lambda_{\rm g},\;\lambda_{\rm e}$ are the corresponding
 eigenvalues. In this case, the ground state is one of the
 N\'eel states and the summation extends over the first
 excited states only. Also, the Fourier transform of $S^x$, i.e., 
 \beqy
S^x(q)=\frac{1}{2\sqrt N}\sum_je^{iqr_j}\left( S^+_j+S^-_j\right). \nonumber
 \eeqy
Similarly, the dynamical spin correlation function perpendicular to
the direction of $h_x$, $S^{yy}(q,\omega)$, is defined by
replacing the superscript $x$ with $y$ in Eq.(\ref{eqsxx}), where
  \beqy
S^y(q)=\frac{1}{2i\sqrt N}\sum_je^{iqr_j}\left( S^+_j-S^-_j\right). \nonumber
 \eeqy
 Since the ground state is the N\'eel state, $S^{xx}(q,\omega)$
 and $S^{yy}(q,\omega)$ directly reflect the wave number
 dependence of the excited states.
\begin{figure}[p]
\begin{center}
\psfrag{sx}{$S^{xx}(q,\omega)$}
\psfrag{sy}{$S^{yy}(q,\omega)$}
\psfrag{k0}{$q\!=\!0$}
\psfrag{k5}{$q\!=\!\frac{\pi}{2}$}
\psfrag{kp}{$q\!=\!\pi$}
\psfrag{w}{$\omega$}
\includegraphics{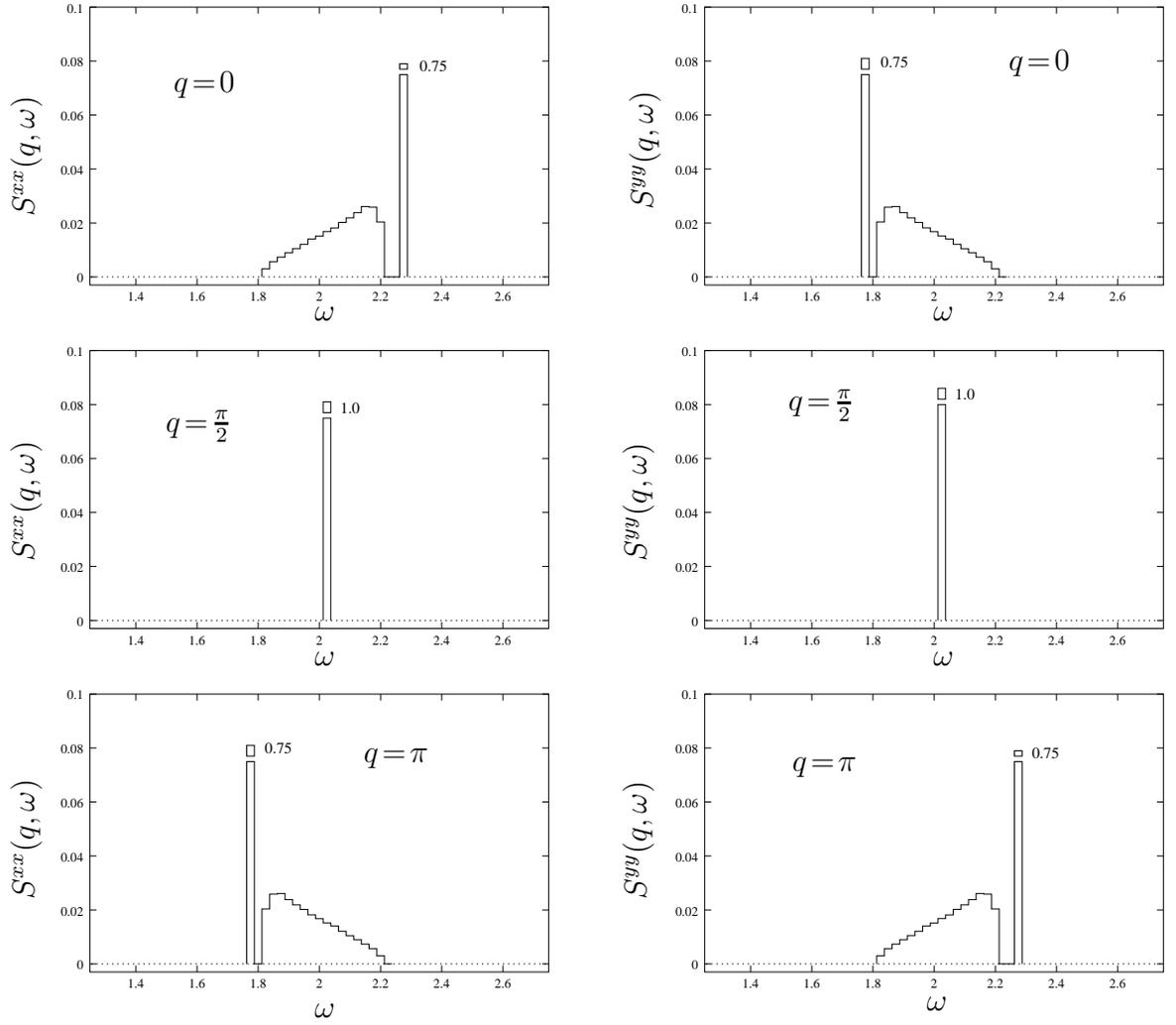}
\label{fig3}
\caption{The functions $S^{xx}(q,\omega)$ and $S^{yy}(q,\omega)$ for $q$ = 0, 
$\frac{\pi}{2}$ and $\pi$ in the histogram with $\Delta \omega$ = 0.025$J$. }
\end{center}
\end{figure}
 With the help of the Eq.(\ref{eqeif}), the dynamical spin
 correlation functions can further be written as \cite{Murao}
 \beqy
S^{xx}(q,\omega)&=& \frac{1}{4}\sum_\alpha |\alpha_1|^2
\delta(\omega-\lambda_\alpha+\lambda_{\rm g}),\nonumber \\
S^{yy}(q,\omega)&=& \frac{1}{4}\sum_\beta |\beta_1|^2
\delta(\omega-\lambda_\beta+\lambda_{\rm g}).
\eeqy
Note that $S^{xx}(q,\omega)$ depends only on $|\alpha_1|^2$ while
 $S^{yy}(q,\omega)$ on $|\beta_1|^2$. Thus, the
 symmetric mode is directly reflected on $S^{xx}(q,\omega)$,
 whereas the antisymmetric mode on $S^{yy}(q,\omega)$. The
 functions $S^{xx}(q,\omega)$ and $S^{yy}(q,\omega)$ for $h_x=0$
 are shown in figure 3. A sharp peak originates
 from the bound state, whareas the broad peak from the
 free DWP states. The intensity of the sharp peak does not
 depend on the number of spins $N$, while the broad peak comprises
 ($N-1$) peaks of which has intensity of the order of $\frac{1}{N}$.
 Note that at the zone-boundary ($q=\frac{\pi}{2}$),
 the width of the continuum vanishes. This is also verified
 in neutron scattering experiments on CsCoCl$_3$ \cite{Yoshizawa}.
\begin{figure}[p]
\begin{center}
\psfrag{w}{$\omega$}
\psfrag{sx}{$S^{xx}(q,\omega)$}
\psfrag{sy}{$S^{yy}(q,\omega)$}
\psfrag{k0}{$q\!=\!0$}
\psfrag{k5}{$q\!=\!\frac{\pi}{2}$}
\psfrag{kp}{$q\!=\!\pi$}
\psfrag{k8}{$q\!=\!0.8\pi$}
\psfrag{k6}{$q\!=\!0.6\pi$}
\psfrag{k4}{$q\!=\!0.4\pi$}
\psfrag{k2}{$q\!=\!0.2\pi$}
\includegraphics{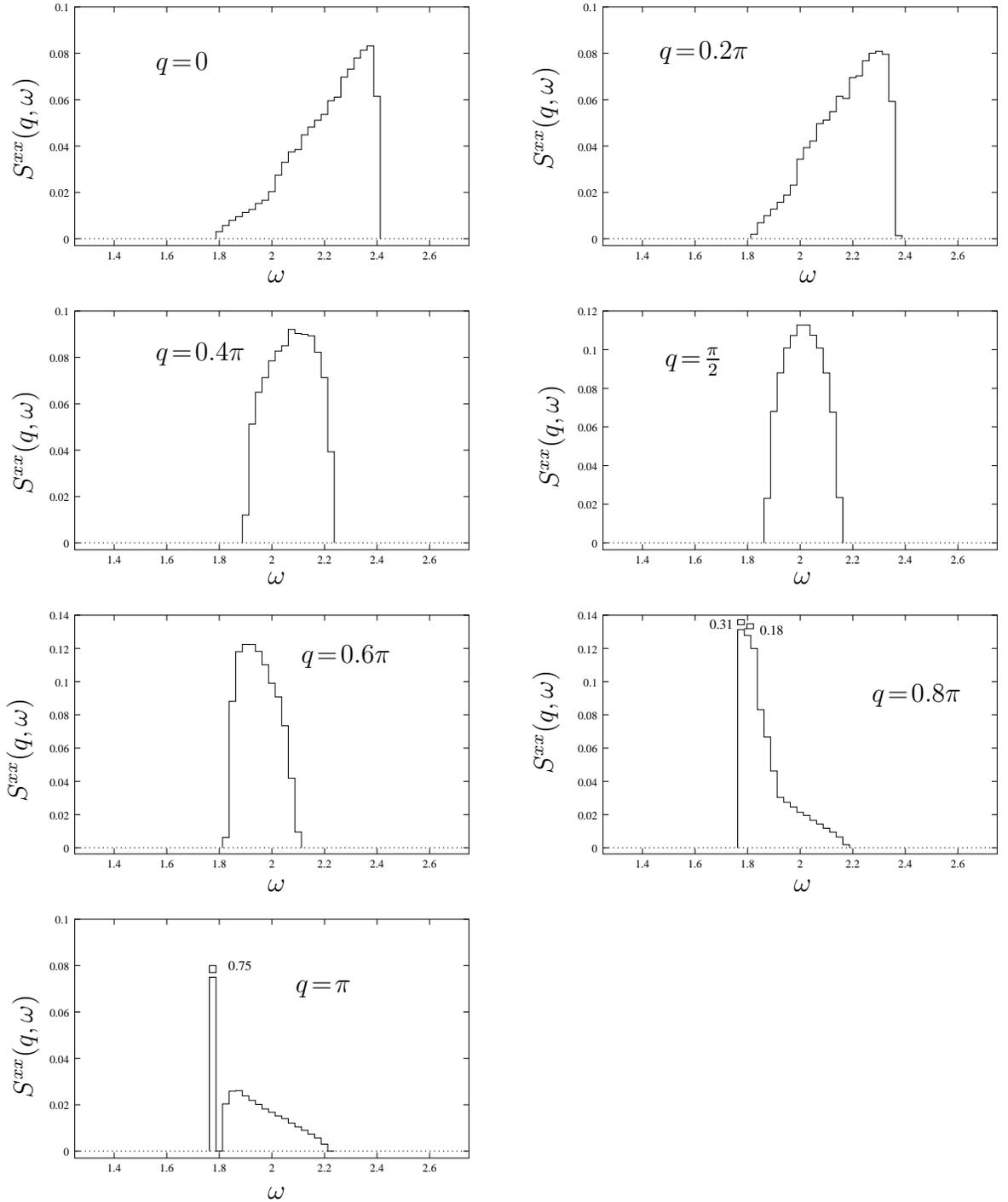}
\label{fig4}
\caption{The function $S^{xx}(q,\omega)$  for different values of $q$ 
and $h_x\!=\!0.1J$. The width of the  histogram is $\Delta \omega$ = 0.025$J$.}
\end{center}
\end{figure}          

\begin{figure}[p]
\begin{center}
\psfrag{w}{$\omega$}
\psfrag{sx}{$S^{xx}(q,\omega)$}
\psfrag{sy}{$S^{yy}(q,\omega)$}
\psfrag{k0}{$q\!=\!0$}
\psfrag{k5}{$q\!=\!\frac{\pi}{2}$}
\psfrag{kp}{$q\!=\!\pi$}
\psfrag{k8}{$q\!=\!0.8\pi$}
\psfrag{k6}{$q\!=\!0.6\pi$}
\psfrag{k4}{$q\!=\!0.4\pi$}
\psfrag{k2}{$q\!=\!0.2\pi$}
\psfrag{h2}{$h_x=0.2J$}
\psfrag{h1}{$h_x=0.1J$}
\psfrag{h5}{$h_x=0.05J$}
\psfrag{h0}{$h_x=0$}         
\label{fig5}
\includegraphics{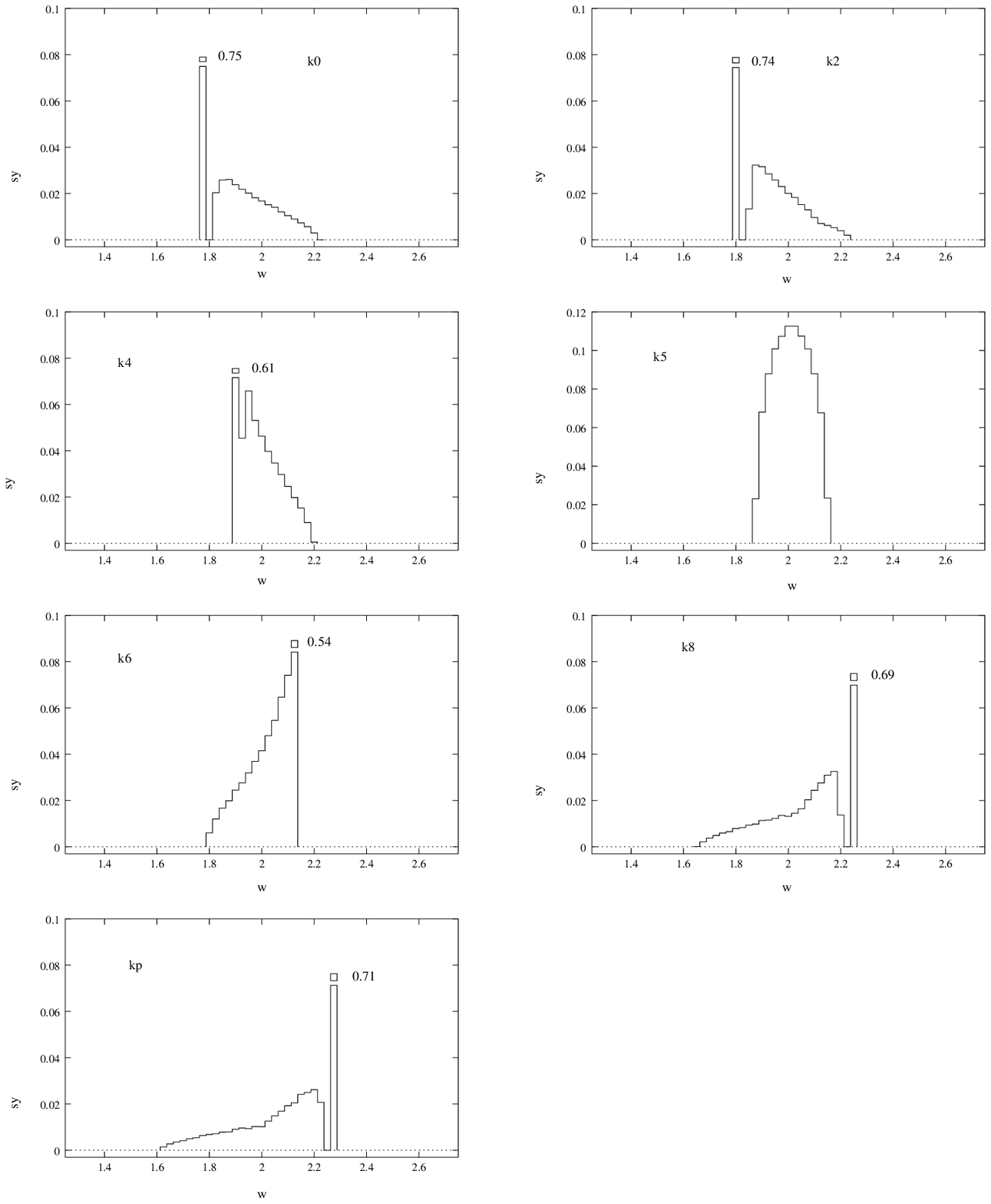}
\caption{The function $S^{yy}(q,\omega)$  for different values of $q$
and $h_x\!=\!0.1J$. The width of the  histogram is $\Delta \omega$ = 0.025$J$.}
\end{center}
\end{figure}

The line shapes of $S^{xx}(q,\omega)$ at $h_x=0.1J$ have been
 plotted in figure 4. The main feature of
 $S^{xx}(q,\omega)$ induced by $h_x$ shows that the line
 shape is purely symmetric at the zone boundary and it is highly
 asymmetric away from the zone boundary. At $q\approx 0$,
 the sharp peak occurs at high energy and the tail at
 lower energy region. For $0<q<\frac{\pi}{2}$, the spectral
 weight concentrates mainly in the middle of the continuum.
 The sharp peak emerges again in the lower energy region for
 $\frac{\pi}{2}<q<\pi$. At $q\approx \pi$, the line shape
 is not affected by $h_x$ as expected from the dispersion
 relation shown in figure 2. Figure 5 
 shows $S^{yy}(q,\omega)$ at $h_x=0.1J$. The line shapes of
 $S^{yy}(q,\omega)$ is again symmetric at $q=\frac{\pi}{2}$.
 The $q$ dependence of $S^{yy}(q,\omega)$ is opposite
 from that of $S^{xx}(q,\omega)$ and it
remains unaffected at $q=0$. For $0<q<\frac{\pi}{2}$,
the sharp peak appears on the lower energy region of the
broad peak. The sharp peak originates from the bound state,
whereas the broad peak from the DWP continuum. For $\frac{\pi}{2}
<q<\pi$, the sharp peak appears on the higher energy side of
the broad peak and the tail is found to enhance towards
the lower energy region. In figures 6 and
7, $S^{xx}(q,\omega)$ and $S^{yy}(q,\omega)$ are
shown for different values of $h_x$, respectively. With the
increase of $h_x$, the features mentioned above are enhanced.
The height of the sharp peak is found to diminish with the
increase of $h_x$. Note that $S^{xx}(q,\omega)$ for
$0<q<\frac{\pi}{2}$ and $S^{yy}(q,\omega)$ for
$\frac{\pi}{2}<q<\pi$ are sensitive on $h_x$ as observed
by Murao $et.\,al.$ \cite{Murao}.
                             
\begin{figure}[p]
\begin{center}
\psfrag{w}{$\omega$}
\psfrag{sx}{$S^{xx}(q,\omega)$}
\psfrag{sy}{$S^{yy}(q,\omega)$}
\psfrag{k0}{$q\!=\!0$}
\psfrag{k5}{$q\!=\!\frac{\pi}{2}$}
\psfrag{kp}{$q\!=\!\pi$}
\psfrag{k8}{$q\!=\!0.8\pi$}
\psfrag{k6}{$q\!=\!0.6\pi$}
\psfrag{k4}{$q\!=\!0.4\pi$}
\psfrag{k2}{$q\!=\!0.2\pi$}
\psfrag{h2}{$h_x\!=\!0.2J$}
\psfrag{h1}{$h_x\!=\!0.1J$}
\psfrag{h5}{$h_x\!=\!0.05J$}
\psfrag{h0}{$h_x\!=\!0$}         
\includegraphics{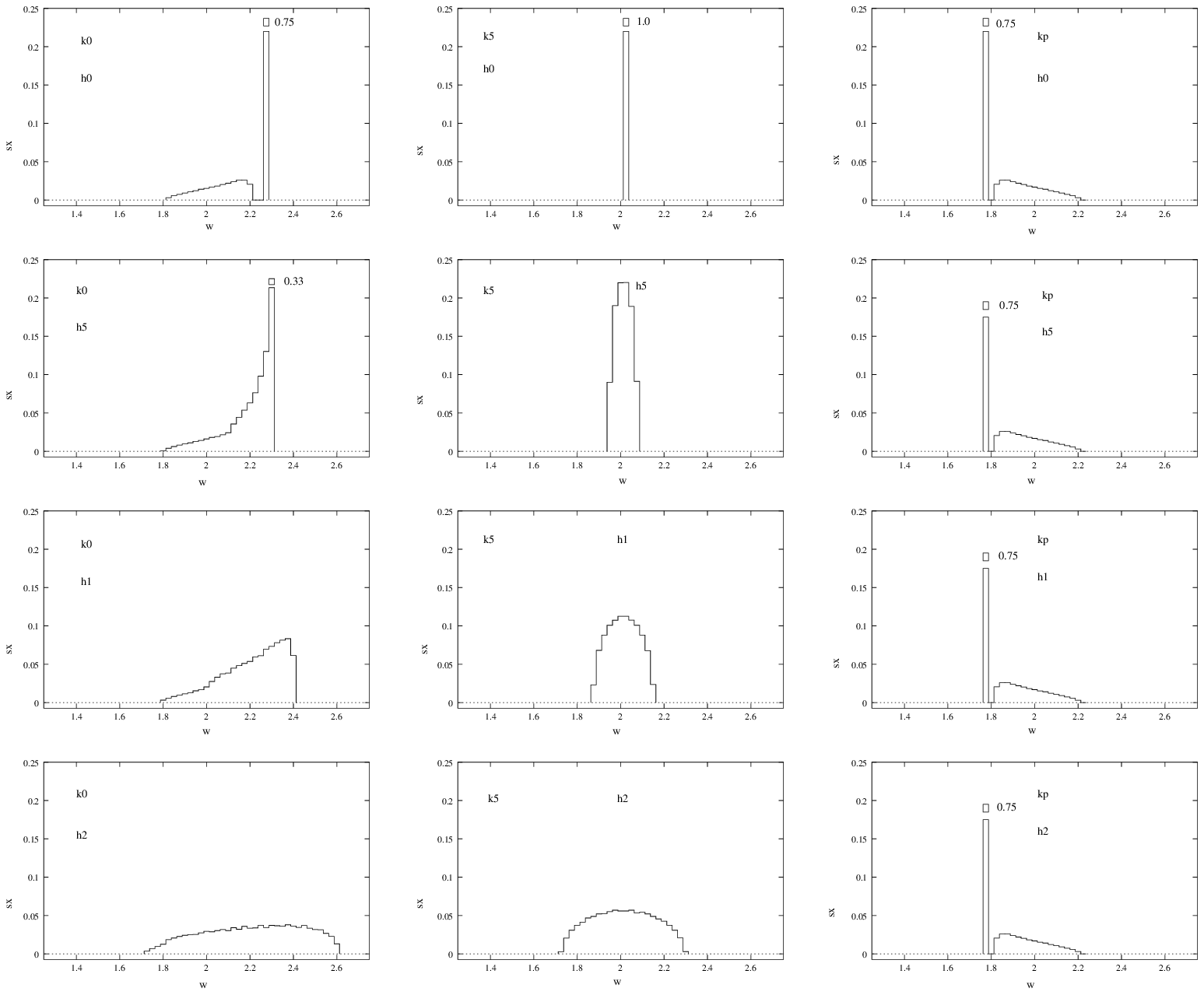}
\label{fig6}
\caption{The function $S^{xx}(q,\omega)$  for different values of $q$
and $h_x$. The width of the  histogram is $\Delta \omega$ = 0.025$J$.}
\end{center}
\end{figure}         
\begin{figure}[p]
\begin{center}
\psfrag{w}{$\omega$}
\psfrag{sx}{$S^{xx}(q,\omega)$}
\psfrag{sy}{$S^{yy}(q,\omega)$}
\psfrag{k0}{$q\!=\!0$}
\psfrag{k5}{$q\!=\!\frac{\pi}{2}$}
\psfrag{kp}{$q\!=\!\pi$}
\psfrag{k8}{$q\!=\!0.8\pi$}
\psfrag{k6}{$q\!=\!0.6\pi$}
\psfrag{k4}{$q\!=\!0.4\pi$}
\psfrag{k2}{$q\!=\!0.2\pi$}
\psfrag{h2}{$h_x\!=\!0.2J$}
\psfrag{h1}{$h_x\!=\!0.1J$}
\psfrag{h5}{$h_x\!=\!0.05J$}
\psfrag{h0}{$h_x\!=\!0$}   
\includegraphics{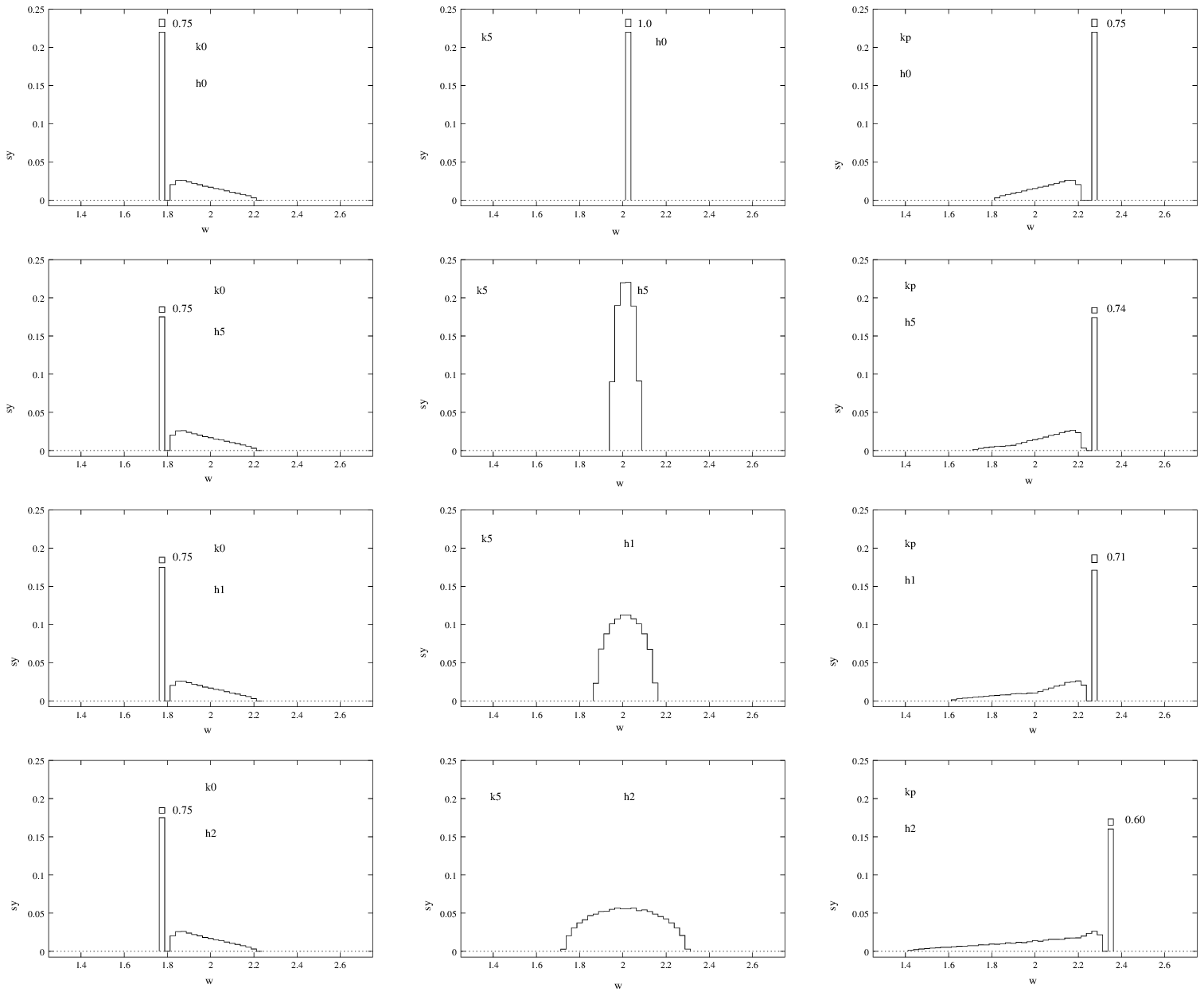}
\label{fig7}
\caption{The function $S^{yy}(q,\omega)$  for different values of $q$
and $h_x$. The width of the  histogram is $\Delta \omega$ = 0.025$J$.}
\end{center}
\end{figure}             

\subsection{Interchain interactions in CsCoCl$_3$}
Now we add the staggered field term to the Hamiltonian in
Eq.(\ref{hxyz}). The full Hamiltonian looks like
 \beqy
H&=&2J\sum_i^N \left[ S_i^zS_{i+1}^z+\frac{\epsilon_1}{2}
\left( S_i^+S_{i+1}^-+S_i^-S_{i+1}^+\right)+
\frac{\epsilon_2}{2}
\left( S_i^+S_{i+1}^++S_i^-S_{i+1}^-\right)
\right] \nonumber \\ && -h_x\sum_i^NS_i^x
-h_{\rm ic}\sum_i^N(-1)^iS_i^z.
\label{hh}
\eeqy
Here, we further consider that the staggered field $h_{\rm ic}$
originates due to the weak interchain interaction. The
interchain interaction has been treated in the
mean-field  approximation. Thus,
\beqy
h_{\rm ic}=2J^\prime\sum_\delta\langle S^z_{i+\delta}\rangle\,,
\label{hic}
\eeqy
where $J^\prime$ is the interchain interaction strength and
$\delta $ is the nearest neighbour on the $ab$ plane.
Following the same technique developed in Section 2, we obtain
$S^{xx}(q,\omega)$ and $S^{yy}(q,\omega)$ numerically.

\begin{figure}[p]
\begin{center}
\psfrag{w}{$\omega$}
\psfrag{sx}{$S^{xx}(q,\omega)$}
\psfrag{sy}{$S^{yy}(q,\omega)$}
\psfrag{k0}{$q\!=\!0$}
\psfrag{k5}{$q\!=\!\frac{\pi}{2}$}
\psfrag{kp}{$q\!=\!\pi$}
\psfrag{k8}{$q\!=\!0.8\pi$}
\psfrag{k6}{$q\!=\!0.6\pi$}
\psfrag{k4}{$q\!=\!0.4\pi$}
\psfrag{k2}{$q\!=\!0.2\pi$}
\psfrag{h0}{$h_{\rm ic}\!=\!0$}
\psfrag{h2}{$h_{\rm ic}\!=\!0.02J$}
\psfrag{h4}{$h_{\rm ic}\!=\!0.04J$}
\psfrag{h6}{$h_{\rm ic}\!=\!0.06J$}    
\includegraphics{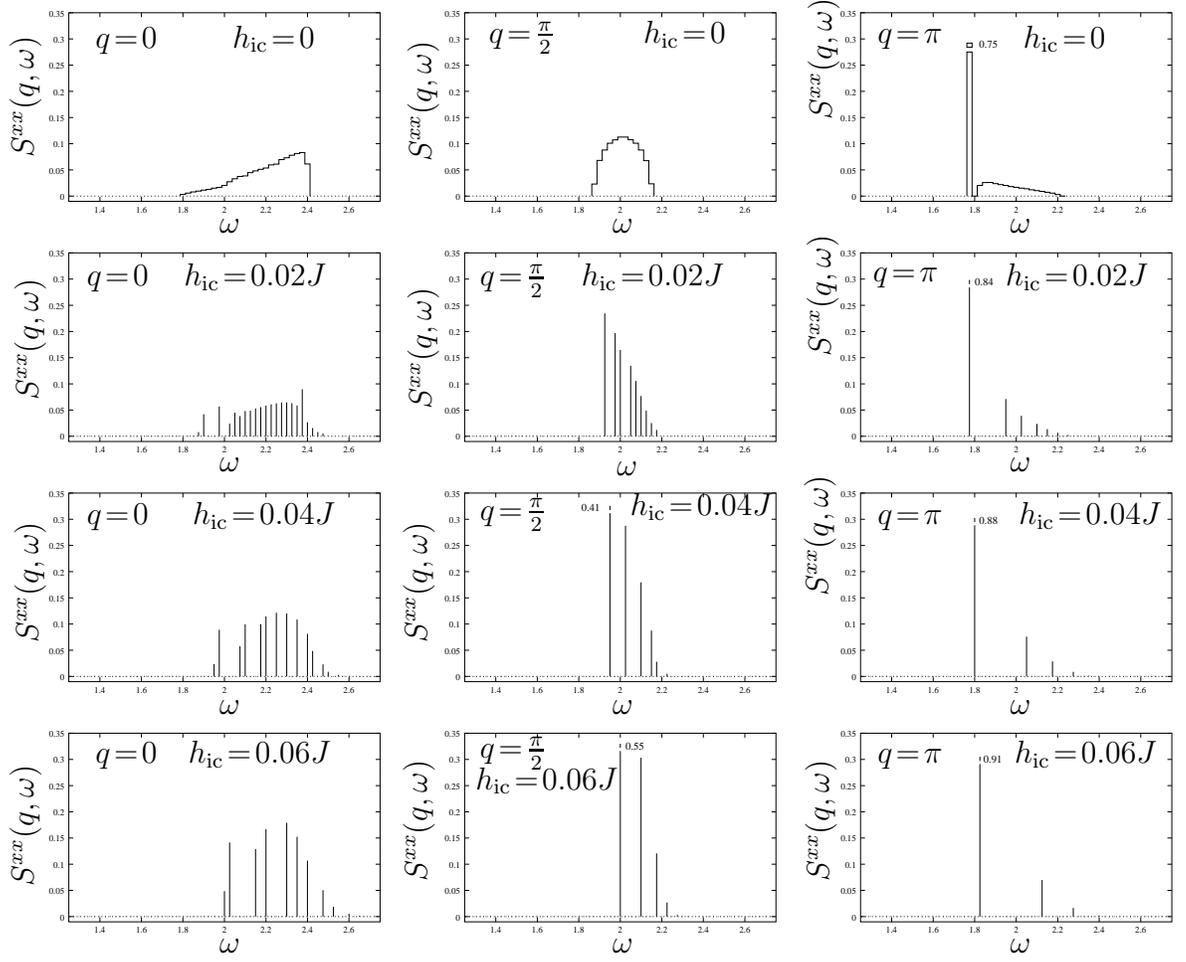}
\label{fig8}
\caption{The function $S^{xx}(q,\omega)$  for different values of $h_{\rm ic}$
and $h_x\!=\!0.1J$. The width of the  histogram is $\Delta \omega$ = 0.025$J$.}
\end{center}
\end{figure}           
\begin{figure}[p]
\begin{center}
\psfrag{w}{$\omega$}
\psfrag{sx}{$S^{xx}(q,\omega)$}
\psfrag{sy}{$S^{yy}(q,\omega)$}
\psfrag{k0}{$q\!=\!0$}
\psfrag{k5}{$q\!=\!\frac{\pi}{2}$}
\psfrag{kp}{$q\!=\!\pi$}
\psfrag{k8}{$q\!=\!0.8\pi$}
\psfrag{k6}{$q\!=\!0.6\pi$}
\psfrag{k4}{$q\!=\!0.4\pi$}
\psfrag{k2}{$q\!=\!0.2\pi$}
\psfrag{m0}{$h_{\rm ic}\!=\!0$}
\psfrag{m2}{$h_{\rm ic}\!=\!0.02J$}
\psfrag{m4}{$h_{\rm ic}\!=\!0.04J$}
\psfrag{m6}{$h_{\rm ic}\!=\!0.06J$}
\includegraphics{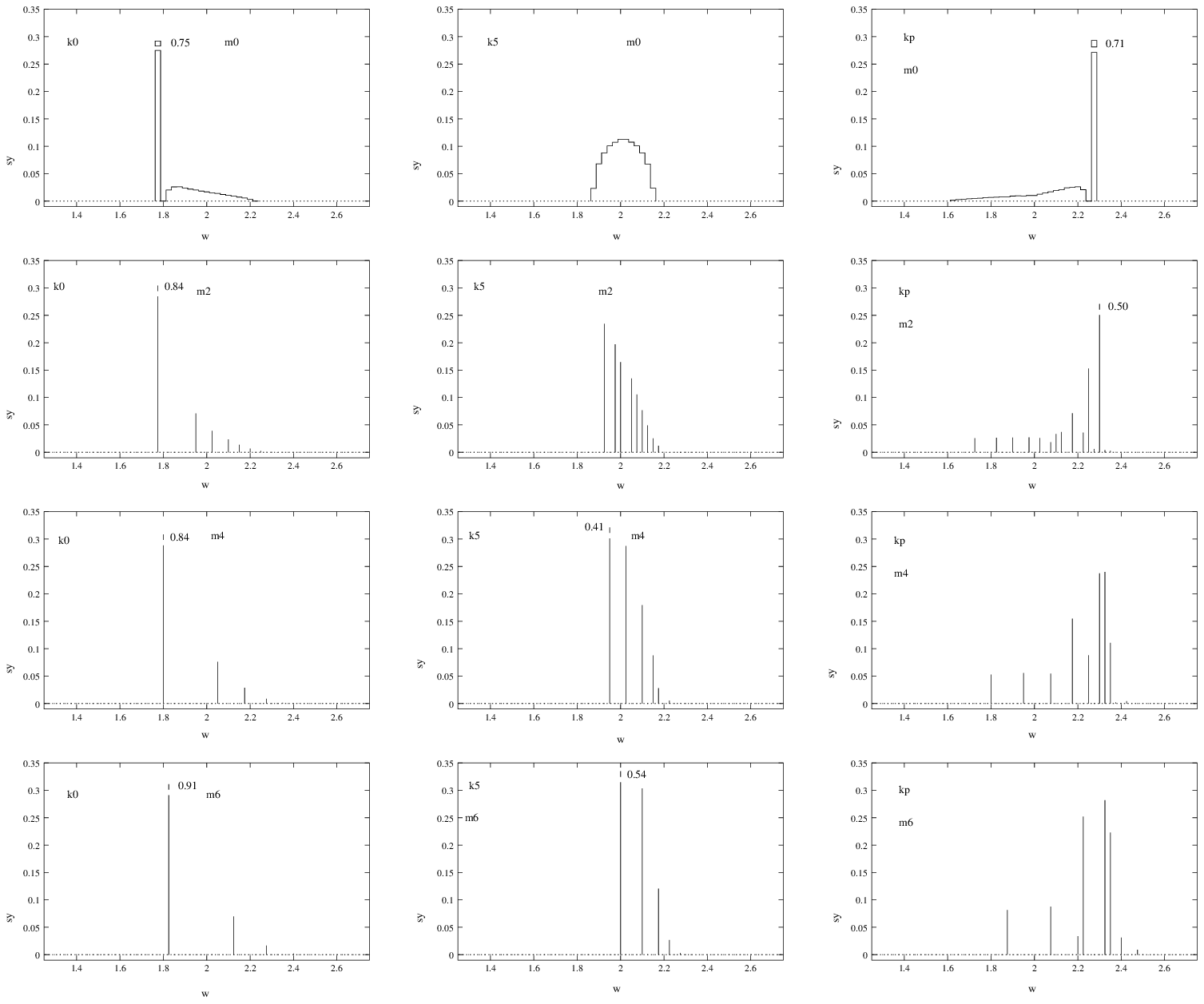}
\label{fig9}
\caption{The function $S^{yy}(q,\omega)$  for different values of $h_{\rm ic}$
and $h_x\!=\!0.1J$. The width of the  histogram is $\Delta \omega$ = 0.025$J$.}
\end{center}
\end{figure}           
In figure 8, $S^{xx}(q,\omega)$ for different
values of $h_{\rm ic}$ are shown. When $h_{\rm ic}\neq 0$,
the broad peak of  $S^{xx}(q,\omega)$ has been splitted into
discrete peaks, which are known as Zeeman ladder \cite{Shiba}.
With the increase of $h_{\rm ic}$, the number of peaks decreases and 
the separation between them becomes wider. For $\frac{\pi}{2}<q<\pi$,
the intensity of the peak at the lowest energy is stronger.
Several peaks having nearly the same intensity are observed
at $q\approx 0$. Figure 9 shows  $S^{yy}(q,\omega)$
for different values of $h_{\rm ic}$.
For $0\leq q \leq \frac{\pi}{2}$, several strong peaks appear in the
lower energy region. Note that the difference between
$S^{xx}(q,\omega)$ and $S^{yy}(q,\omega)$ are remarkable at
$q=\pi$. In  $S^{yy}(q,\omega)$ for $q=\pi$, several strong peaks
appear in the higher energy region and weak peaks in the lower
energy region, and their intensities vary irregularly with the
increase in $h_{\rm ic}$. Thus the combined effect of $h_x$ and
the interchain interactions is also found in
$S^{xx}(q,\omega)$ and $S^{yy}(q,\omega)$ for
$q\geq \frac{\pi}{2}$ as reported in the Ref.\cite{Murao}.

\begin{figure}[t]
\begin{center}
\psfrag{a}{(a)}
\psfrag{b}{(b)}
\includegraphics{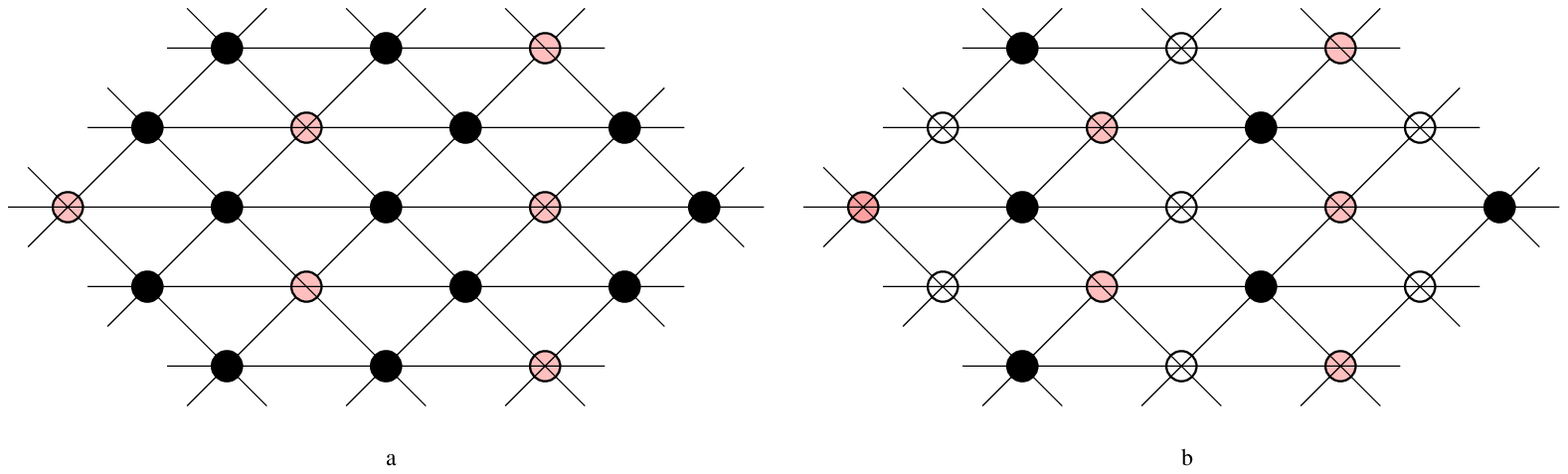}
\label{fig10}
\caption{Magnetic ordering in $ab$ plane below $T_{\rm N_1}$: (a)
Ferrimagnetic structure for $T<T_{\rm N_2}$ (F-phase). (b) Structure for
 $T_{\rm N_2}<T<T_{\rm N_1}$ (A-phase).  Chains marked with open circles
are disordered.}
\end{center}
\end {figure}

In CsCoCl$_3$, the magnetic Co$^{+2}$ ions are surrounded by
trigonally distorted octahedra of Cl$^-$ ions and form
chains along the $c$-axis with successive octahedra sharing
a common face. The CsCl$^-_3$ chains are arranged in a triangular
array. Since the exchange coupling between chains is
antiferromagnetic, the triangular array forms a frustrated
system. Thus there is no possibility of a perfectly regular AFM
ordered state. Two different three-dimensionally ordered phases
occur in CsCoCl$_3$. First, below $T_{N_1}\sim$ 21K
partially disordered AFM phase (A) is formed in which one-third
of the chains are paramagnetic. A phase change takes place below
$T_{N_2}\sim$ 10-14K, to a ferrimagnetic phase (F), in which
the paramagnetic chains align in the same direction, so that
two-thirds of the chains are aligned in one direction and
one-third in the opposite direction (Fig. 10) \cite{Mekata}.

\begin{figure}[p]
\begin{center}
\psfrag{w}{$\omega$}
\psfrag{sx}{$S^{xx}(q,\omega)$}
\psfrag{sy}{$S^{yy}(q,\omega)$}
\psfrag{k0}{$q\!=\!0$}
\psfrag{k5}{$q\!=\!\frac{\pi}{2}$}
\psfrag{kp}{$q\!=\!\pi$}
\psfrag{k8}{$q\!=\!0.8\pi$}
\psfrag{k6}{$q\!=\!0.6\pi$}
\psfrag{k4}{$q\!=\!0.4\pi$}
\psfrag{k2}{$q\!=\!0.2\pi$}
\psfrag{h2}{$h_x\!=\!0.2J$}
\psfrag{h1}{$h_x\!=\!0.1J$}
\psfrag{h5}{$h_x\!=\!0.05J$}
\psfrag{h0}{$h_x\!=\!0$} 
\psfrag{a}{}
\psfrag{b}{}
\includegraphics{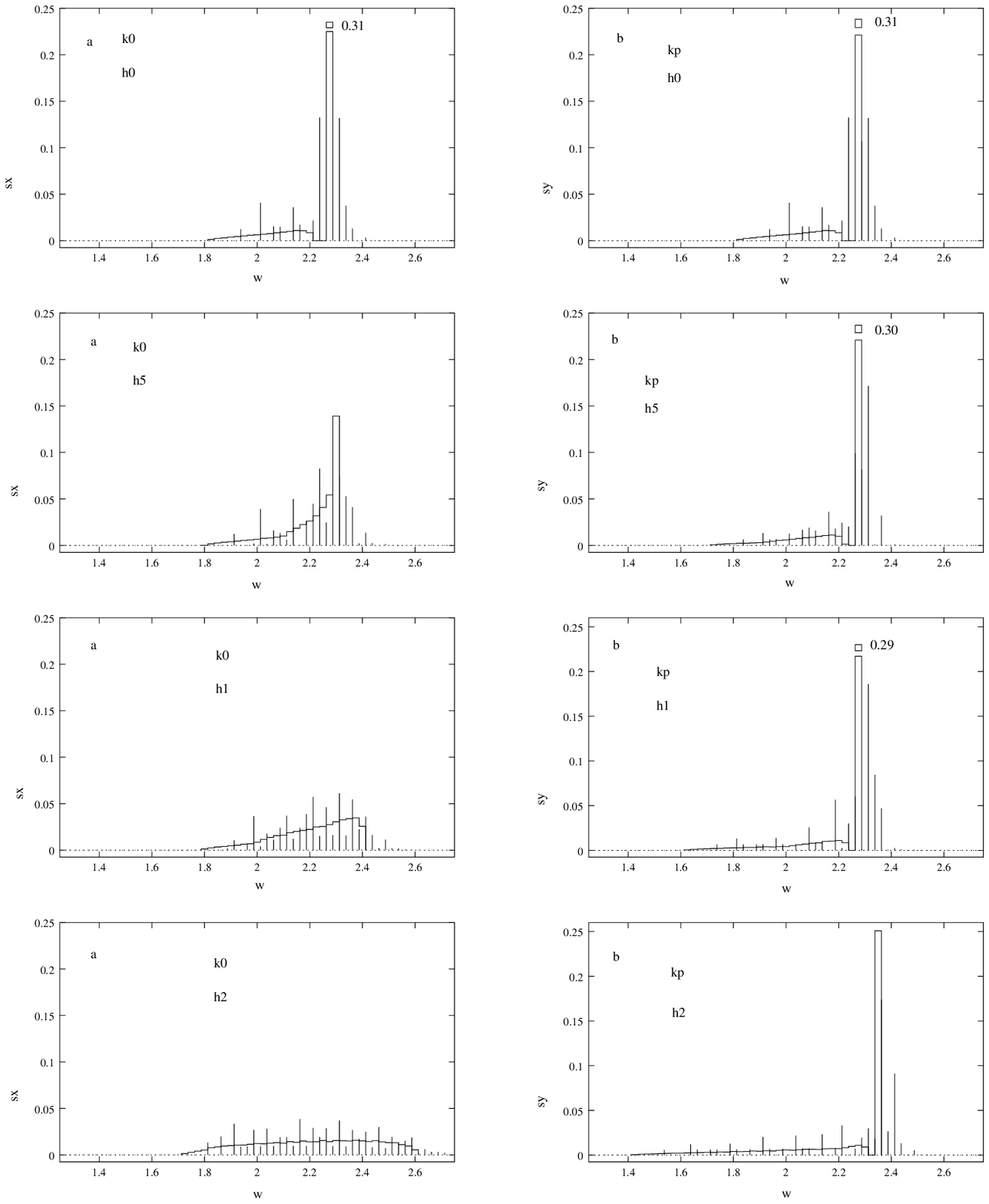}
\label{fig11}
\caption{The function $S^{xx}(q,\omega)$  for $q\!=\!0$ and  
$S^{yy}(q,\omega)$  for $q\!=\!\pi$ in the ferrimagnetic phase. 
 The width of the  histogram is $\Delta \omega$ = 0.025$J$.}
\end{center}
\end{figure}          
 
\begin{figure}[p]
\begin{center}
\psfrag{w}{$\omega$}
\psfrag{sx}{$S^{xx}(q,\omega)$}
\psfrag{sy}{$S^{yy}(q,\omega)$}
\psfrag{k0}{$q\!=\!0$}
\psfrag{k5}{$q\!=\!\frac{\pi}{2}$}
\psfrag{kp}{$q\!=\!\pi$}
\psfrag{k8}{$q\!=\!0.8\pi$}
\psfrag{k6}{$q\!=\!0.6\pi$}
\psfrag{k4}{$q\!=\!0.4\pi$}
\psfrag{k2}{$q\!=\!0.2\pi$}
\psfrag{h2}{$h_x\!=\!0.2J$}
\psfrag{h1}{$h_x\!=\!0.1J$}
\psfrag{h5}{$h_x\!=\!0.05J$}
\psfrag{h0}{$h_x\!=\!0$}  
\psfrag{a}{}
\psfrag{b}{}
\includegraphics{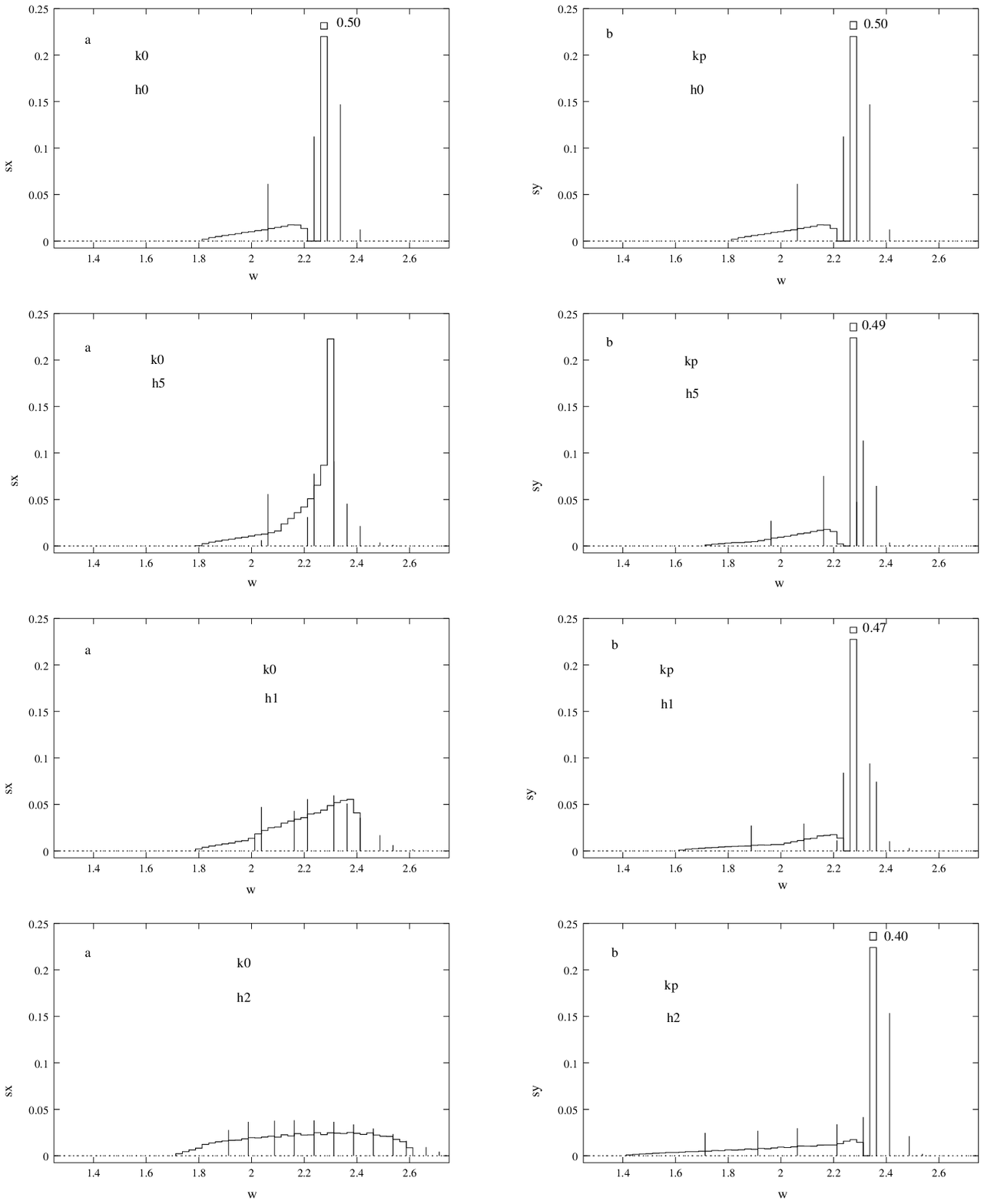}
\label{fig12}
\caption{The function $S^{xx}(q,\omega)$  for $q\!=\!0$ and
$S^{yy}(q,\omega)$  for $q\!=\!\pi$ in the partially disordered phase.
 The width of the  histogram is $\Delta \omega$ = 0.025$J$.}
\end{center}
\end{figure}           

It is obvious from the Eq.(\ref{hic}), that the staggered
field $h_{\rm ic}$ could take one of the two possible values of
$h_{\rm ic}= 6J^\prime$ and 0. Similarly, in the partially disordered
phase, $h_{\rm ic}$ takes one of the four values of
$h_{\rm ic}= 6J^\prime ,\,4J^\prime ,\,2J^\prime$ and 0.
The functions $S^{xx}(q,\omega)$ and $S^{yy}(q,\omega)$ of
CsCoCl$_3$ are obtained by summing up those of the individual
chains. In the ferrimagnetic phase, ratio of the number
of chains with $h_{\rm ic}= 6J^\prime$ and 0 are estimated
as $1:2$. In the partially disordered
phase, that with  $h_{\rm ic}= 6J^\prime ,\,4J^\prime ,
\,2J^\prime$ and 0 are estimated as $1:3:3:5$ \cite{Murao}.  
Figure 11 shows $S^{xx}(q,\omega)$
for $q=0$ and $S^{yy}(q,\omega)$
for $q=\pi$ in the ferrimagnetic phase. The line shape of
$S^{xx}(q,\omega)$  is affected strongly by the $h_x$.
As $h_x$ increases, the difference in intensities among
various discrete peaks reduce markedly and the sharp peak 
disappears. On the other hand,
$S^{yy}(q,\omega)$  
is affected slightly by $h_x$.
The spectral weight enhances towards the lower energy region
with the increase in $h_x$, and the intensity distribution varies
irregularly. $S^{xx}(q,\omega)$  for $q=0$ and
$S^{yy}(q,\omega)$ for $q=\pi$ obtained in the partially
disordered phase are shown in figure 12. 
$S^{xx}(q,\omega)$ is strongly affected by $h_x$ 
while $S^{yy}(q,\omega)$ is much less. $S^{xx}(q,\omega)$ 
for $q=\pi$ and $S^{yy}(q,\omega)$ 
for $q=0$ are not affected by $h_x$ both in the ferrimagnetic and the 
partially disordered phases. 
\section{Discussion of results}
 We have studied the effect of the transverse magnetic field
 $h_x$ on dynamical properties of one dimensional
 fully anisotropic Ising-like antiferromagnet at low temperatures.
 We have shown using this Hamiltonian that some of the
 results obtained by Murao $et.\,al.$ \cite{Murao}, in which
 a FM NNN interaction is assumed besides the usual AFM NN
 interaction, can be qualitatively reproduced. These include
 the formation of DWP bound states, two types of excited
 modes which are symmetric and antisymmetric with respect to
 the states with $S^z_{\rm T}=1$ and $-1$ and an
 asymmetry in the line shape of the correlation functions
$S^{xx}(q,\omega)$ and $S^{yy}(q,\omega)$. In order to obtain
the asymmetry in the line shape of
$S^{xx}(q,\omega)$ and $S^{yy}(q,\omega)$, if a FM NNN exchange
of magnitude $|J^\prime|\sim 0.1 |J|$ is required, which
considering that the NNN exchange is through two nonmagnetic
ligands would seem to be unphysically large \cite{Goff}.
On the other hand, our model could explain all these
charateristics with the usual NN AFM exchange interactions.
There are, however, a number of differences. Murao $et.\,al.$
\cite{Murao} observed a single bound state branch which is
symmetric with respect to the zone boundary, whereas in the
present study, different bound state branches are obtained
for symmetric and antisymmetric modes which are asymmetric
with respect to the zone boundary. No experimental evidence
are as yet available on the effect of bound states on the
thermodynamic and dynamic properties of the compounds
CsCoCl$_3$ and CsCoBr$_3$. The symmetric modes contribute
to $S^{xx}(q,\omega)$, whereas antisymmetric modes contribute to
$S^{yy}(q,\omega)$. Both $S^{xx}(q,\omega)$ and
$S^{yy}(q,\omega)$ have symmetry at the zone boundary even in
the presence of either both $h_x$ and $h_{\rm ic}$
or any one of them. This symmetry is totally lost away from
the zone boundary. This theory is also valid to analysis of
spin dynamics of  CsCoCl$_3$ and CsCoBr$_3$ at finite
temperatures, because the correlation length of the spin
along the chain is very large even in the paramagnetic phase
($T\approx 21$K). Thus the effects of $h_x$ on
$S^{xx}(q,\omega)$ and $S^{yy}(q,\omega)$ discussed here are
yet to be observed in a real system.

Apart from relevance to experimental systems such as
 CsCoCl$_3$ and CsCoBr$_3$, the present study
 is intended to provide insights about the spin dynamics
 of fully anisotropic Ising-like AFM system in the
 presence of transverse magnetic field $h_x$. The ground state
 energy and low lying excitation spectrum of the fully
 anisotropic Hamiltonian are known exactly because of the
 mapping between the fully anisotropic Hamiltonain and
 the exactly solvable eight vertex model \cite{Johnson,Baxter}.
  Our calculations provide us with
 some physical insights about spin dynamics in
 Ising-like  fully anisotropic AFM system in the
 presence of transverse magnetic field.

\end{document}